\documentclass[12pt,preprint]{aastex}

\begin{document}

\title{Possible Infrared Counterparts to the Soft Gamma-Ray Repeater
SGR 1806-20}

\author{S.S. Eikenberry\altaffilmark{1}, 
M.A. Garske\altaffilmark{1,2}, D. Hu\altaffilmark{1},
M.A. Jackson\altaffilmark{1}, S.G. Patel\altaffilmark{1},
D.J. Barry\altaffilmark{1}, M.R. Colonno\altaffilmark{1},
J.R. Houck\altaffilmark{1}}

\altaffiltext{1}{Astronomy Department, Cornell University, Ithaca, NY 14853}

\altaffiltext{2}{Physics Department, Northwest Nazarene University, Nampa, ID  83686}

\begin{abstract}

	We report the discovery of possible infrared counterparts to
SGR 1806-20.  We use archival {\it Chandra} observations to determine
the location of SGR 1806-20 to $<1 \arcsec$ accuracy.  We then locate
2 infrared objects within this error circle in K-band ($2.2 \mu$m)
images of this field.  Based on the X-ray absorption towards SGR
1806-20 and the extinction towards the nearby star cluster, we discuss
the likelihood of association for the possible counterparts, and the
implications for SGR 1806-20's physical properties and origins.

\end{abstract}


\section{Introduction}

	The soft gamma-ray repeaters (SGRs) are unusual objects, even
for the rarefied field of compact object astrophysics, with only 4
definitely identified over the last 30 years.  As a class, they are
defined by their repeated super-Eddington soft gamma-ray bursts (peak
photon energy $\sim 30$ keV, versus $\sim 300$ keV for ``classical''
gamma-ray bursters), with occasional outbursts {\it greatly} exceeding
the Eddington luminosity for a $\sim 1 M_{\odot}$ object.  During a
giant outburst, SGR 0526-66 showed 8-second pulsations \citep{Mazets},
and SGR 1900+14 showed 5-second pulsations during a similar outburst
some 20 years later \citep{Hurley99a}.  Both SGR 1900+14 and SGR
1806-20 show X-ray pulsations (5-s and 7.5-s, respectively) in
quiescence (\citet{Chryssa}; \citet{Hurley99b}).  Based on their spin
periods and period derivatives, SGRs appear similar to the so-called
Anomalous X-ray Pulsars (AXPs) \citep{Stella}.  Currently, the most
widely investigated model for both SGRs and AXPs is the so-called
``magnetar'' model, in which a highly-magnetized young neutron star
has X-ray emission powered by the decay of its magnetic field, and the
soft gamma-ray bursts are powered by magnetic realignment in the
neutron star crust (\citet{DT92}; \citet{DT93}).  Other models have
also been proposed (e.g. \citet{Marsden}).

	One motivation for considering SGRs as young neutron stars has
been their apparent association with young supernova remnants,
particularly the association of SGR 1806-20 with the radio nebula
G10.0-0.3 (\citet{Vasisht}; \citet{KF93}).  At one time, SGR 1806-20
was thought to be associated with a luminous blue variable (LBV) star
which lies at the time-variable (in both flux and morphology) core of
this nebula.  However, the recent Inter-Planetary Network (IPN)
localization of SGR 1806-20 provides a position inconsistent with that
of the LBV star and radio core \citep{Hurley99c}.  Furthermore,
\citep{Gaensler} argue that G10.0-0.3 is not a supernova remnant at
all; rather it may be powered by the tremendous wind of the LBV star
\citep{Hurley99c}.  Infrared observations of the field of SGR 1806-20
reveal that the LBV star is not alone, but appears to be part of a
cluster of embedded, hot, luminous stars \citep{Fuchs}, and the IPN
position for SGR 1806-20 is consistent with membership in that
cluster.  Recently, \citet{LBV} have used near-infrared photometry and
spectroscopy to conclude that this cluster contains what may be the
most luminous star in the Galaxy (the LBV star), at least one
Wolf-Rayet star of type WCL, and at least two blue ``hypergiants'' of
luminosity class Ia+.  These properties make the cluster resemble a
somewhat smaller and older version of the ``super'' star cluster R136
\citep{Massey}, making the potential association with SGR 1806-20 even
more intriguing.

	We report here X-ray observations of SGR 1806-20 with the {\it
Chandra X-ray Observatory} \citep{Marty}, as well as near-infrared
observations of the field with the Cerro Tololo Inter-American
Observatory (CTIO) 4-meter telescope and the Hartung-Boothroyd
Observatory 0.65-m telescope.  Using {\it Chandra} to localize SGR
1806-20 to an accuracy $<1 \arcsec$, we then can identify possible
infrared (IR) counterparts to SGR 1806-20.  In Section 2, we describe
the observations and data analysis.  In Section 3, we discuss the
results and their implications for SGR 1806-20 and other soft
gamma-ray repeaters, and in Section 4 we present our conclusions.

\section{Observations \& Analysis}

\subsection{{\it Chandra} Observations}

	We analyzed an archival {\it Chandra} observation of SGR
1806-20 taken on 24 July 2000 UT.  The observation duration was 4900
seconds, with the aim point on the back-illuminated ACIS-S3 CCD chip.
Because the time-resolution was 3.24-s (full-frame mode), sources as
bright as SGR 1806-20 suffered from significant photon pileup in the
core ($\sim 30-40 \%$ here).  This effect limits spectroscopic
accuracy, and also prevents detailed analysis of the spatial profile
near the (suppressed) image core, but does not greatly impair the
source localization.  We used the {\it CIAO 2.1.3} and {\it SHERPA
2.1.2} software packages for {\it Chandra} to perform the analyses of
the data.

	We present a section of the Chandra image of this field in
Figure 1, revealing a single bright source near the expected position
of SGR 1806-20.  Within the limits imposed by pileup, the source
appears to be unresolved (Gaussian $ \sigma \ = \ 0.8 \ {\rm pixels} \
= \ 0.4 \arcsec$), and has a centroid of $\alpha_{2000} = {\rm 18h \
08m \ 39.32s}$ and $\delta_{2000} = {-20 \degr \ 24 \arcmin \ 39.8
\arcsec}$ based on the {\it Chandra} aspect solution.  The uncertainty
in this position is dominated by the systematic offsets in the {\it
Chandra} aspect solution, which are typically $0.5 - 1.0 \arcsec$.
\citet{Kaplan} have used deeper {\it Chandra} observations of this
field to identify several X-ray counterparts of nearby USNO-A2.0
stars, and based on them find a corrected position of $\alpha_{2000} =
{\rm 18h \ 08m \ 39.32s}$ and $\delta_{2000} = {-20 \degr \ 24 \arcmin
\ 39.5 \arcsec}$ , with uncertainties in each coordinate of $\pm 0.3
\arcsec$, in good agreement with our position above.  Thus, we adopt
the more accurate position of \citet{Kaplan} from here on.

	Despite the limitations imposed by the pulse pileup, we
attempted crude spectral analyses on the {\it Chandra} observation.
We extracted a background-subtracted spectrum for the point source and
then modelled it as an absorbed power-law, obtaining a best-fit with
$\chi^2_{\nu} = 0.88$ for 41 degrees of freedom.  This produced an
absorption estimate of $n_H = 4.5 (\pm 1.0) \ \times 10^{22} \ {\rm
cm^{-2}}$ and a flux of $\sim 6 \times 10^{-12} \ {\rm erg \ cm^{-2} \
s^{-1}}$ in the 2-10 keV band -- both in good agreement with the {\it
BeppoSAX} observations of \citet{Mereg}.

\subsection{Infrared Observations}

	We obtained near-infrared images of the field of SGR 1806-20
using the Ohio State InfraRed Imaging Spectrograph (OSIRIS) intrument
\citep{Depoy} and f/14 tip-tilt secondary on the CTIO 4-meter
telescope on 6 July 2001 UT.  We note that no IPN bursts had been
observed from SGR 1806-20 for 2 weeks prior to this date.  We used the
f/7 camera of OSIRIS, providing a plate scale of $0.161 \arcsec$ per
pixel and an unvignetted field of view of $\sim 100 \arcsec$ on a
side.  We observed the field in the J, H, and K bands (central
wavelengths of $1.25 \mu$m, $1.65 \mu$m, and $2.2 \mu$m,
respectively).  For each band, we obtained 9 separate images in a $3
\times 3$ grid pattern offset by $\sim 10 \arcsec$.  The integration
times for each image were 3.24, 3.24, and 10 seconds for J,H, and K
bands respectively.  We then subtracted dark frames from each image,
divided the result by its own median, and then median-combined the
resulting images into a normalized sky frame.  We subtracted the dark
frame and a scaled version of the sky frame from each of the 9 images,
and divided the result by a dome flat image.  We shifted each of the 9
frames to a common reference position and averaged them to give a
final image in each band.

	Due to the presence of variable thin clouds during the CTIO
observations, we photometrically post-calibrated these images using
observations of stars in the field of SGR 1806-20 and infrared
standard stars on July 27, 2001 with the Hartung-Boothroyd Observatory
0.65-meter telescope and its infrared array camera \citep{Houck}.  We
took 7 images of the field of SGR 1806-20 in each band, with offsets
of $\sim 15 \arcsec$ between images.  We then processed the images in
each band as described for the CTIO data above.  We repeated this
procedure on sets of 7 images of the UKIRT standard FS26 in each band.
We extracted the flux in $\rm ADU / s$ from each processed image of
FS26 individually, using the average as the best estimate of the flux,
and the standard deviation as the $1 \sigma$ uncertainty.  We then
used the known magnitudes of this star to calibrate similarly-derived
flux measurements and uncertainties for several bright stars in the
field of SGR 1806-20.  We used a photometric airmass solution derived
from measurements of stars observed for another program over a range
of airmasses throughout the night.

	For astrometric calibration, we had several USNO-A2.0 stars
present in the J-band image.  We chose 8 of the brightest stars
(Figure 2a), and peformed a least-squares linear fit to determine the
best astrometric solution.  We found an RMS of $< 0.15 \arcsec$ in
each coordinate, including the USNO-A2.0 relative uncertainties which
can be as large as $0.25 \arcsec$.  We then used the centroids of
several IR-bright stars to transfer this coordinate system onto the H-
and K-band images (Figure 2b,c).  We found this approach to provide
superior results to simply applying an astrometric solution to the
USNO stars visible in the H- and K-band images -- reddened stars
greatly increase crowding in these images and thus increase
contamination of the centroids of the USNO stars.  The transfer of
coordinates produced an additional RMS uncertainty of $<0.08 \arcsec$
in each coordinate for the H- and K-band images.  Since the X-ray
position for SGR 1806-20 of \citet{Kaplan} also derives from USNO-A2.0
stars, the two solutions should be directly comparable.  Thus, the
90\% positional uncertainty we can provide for the location of SGR
1806-20 on the infrared images is $0.7 \arcsec$.  We present J-, H-,
and K-band images of the region near SGR 1806-20 in Figure 3, along
with the 90\% positional error circle.  We present photometry of the
labelled stars in Table 1.

\section{Discussion}

\subsection{The Distance to SGR 1806-20}

	As noted above, SGR 1806-20 lies in the direction of an
unusual embedded cluster of massive, luminous young stars
(\citet{Fuchs}; \citet{LBV}), with a distance of $14.5 \pm 1.4$ kpc
and a reddening of $A_V = 29 \pm 2$ mag (\citet{Corbel}; \citet{LBV}).
The LBV star is perhaps the most luminous star known at $L > 6 \times
10^6 \ L_{\odot}$, and the Wolf-Rayet and blue hypergiant stars are
also certain to be very massive \citep{LBV}.  The fact that stars D
and E in Figure 3 have $J-K = 5.0$ mag indicates that they are members
of this cluster ($E_{J-K} = 5.0$ mag for $A_V = 29$ mag), and thus
that SGR 1806-20 lies within the radial extent of the cluster on the
sky.  Furthermore, the X-ray absorption towards SGR 1806-20 is $\sim
5-6 \times 10^{22} \ {\rm cm^{-2}}$ (see above, and \citet{Mereg}),
which is consistent with the extinction towards the cluster.  Thus, it
seems likely that SGR 1806-20 is also a member of this massive star
cluster at a distance of $14.5 \pm 1.4$ kpc.  It is very interesting
to note that the soft gamma-ray repeater SGR 1900+14 is also very near
to a cluster of embedded luminous stars \citep{Vrba}.  The projected
distance from the center of this cluster to the location of SGR
1900+14 is $<1$ pc.

\subsection{The IR Counterparts}

	The IR colors (and upper limits) of the two candidate
counterparts to SGR 1806-20 are consistent with both of them being
stellar members of the star cluster ($J-K = 5.0$ mag, $H-K = 2.0$
mag).  Note that the brighter star just outside the error circle
(``C'' in Figure 3) appears to be a foreground star ($J-K = 3$ mag),
confirming that it is not a likely counterpart to the SGR.  However,
as can been seen in Figure 3, the field of SGR 1806-20 is highly
crowded in the IR by both foreground/background objects and cluster
members, and the simple fact that two stars lie within the 90\%
confidence error circle and another just outside it shows that the
probability of a chance coincidence of unrelated IR objects is high.
Thus, we cannot conclude definitely that {\it either} of the possible
counterparts is actually related to SGR 1806-20.  Monitoring for
variability may resolve this issue in the future.

	If both candidates are in fact members of the cluster, we can
estimate their absolute magnitudes to be $M_K = -2.3$ mag (A) and $M_K
= -0.4$ mag (B).  For Star A, this is consistent with stars of
luminosity matching a B1V or K3III star, and is inconsistent with any
stars of luminosity class I.  After correcting for the $H-K = 2$ mag
differential extinction toward the cluster, the intrinsic color of
$(H-K)_{intrins} = 0.8 \pm 1.2$ mag is essentially consistent with all
stellar spectra earlier than late M, and does not significantly
constrain the classification.  For Star B, the absolute magnitude is
consistent with stars of luminosity matching a B8V star, and is
inconsistent with any stars of luminosity class III or higher.  It is
important to note that no observations have yet probed the
distribution of stars in the cluster with masses below that of late B
main sequence stars.  Thus, it is possible that there are even further
stars within the error circle at significantly lower mass/luminosity,
unless the cluster mass distribution shows a sharp lower cutoff.

\subsection{The SGR Progenitor Star}

	One particularly intriguing aspect of the association between
SGR 1806-20 and the star cluster is that a neutron star progenitor
went supernova {\it before} the stars currently observed in the
cluster.  Since more massive stars evolve to the supernova stage more
rapidly, {\it if} the progenitor of SGR 1806-20 formed at the same
time as the currently observed massive stars, its mass must have been
greater.  However, the mass estimate for the LBV is $>200 M_{\odot}$
\citep{LBV}, and several of the other stars are likely to have masses
in the range of $\sim 50-100 M_{\odot}$ \citep{LBV}.  While recent
theories have predicted that very massive stars may produce neutron
star remnants due to the effect of envelope loss via dense stellar
winds, the upper limit on their masses is $\sim 80 M_{\odot}$, with
higher mass stars producing massive black hole remnants.  Thus, it
seems unlikely that the SGR 1806-20 progenitor formed at the same time
as the currently observed massive stars in the cluster.

	Alternately, SGR 1806-20 may have formed prior to these
stars -- in fact, \citet{Kaplan} suggest that the supernova event that
produced SGR 1806-20 may have triggered the star formation activity
that produced the massive stars in the cluster.  However, the massive
stars we observe in the cluster are evolved, with ages of $\sim 10^6$
yr to reach the LBV and Wolf-Rayet stages of their lives.  If the SGR
1806-20 supernova event led to the birth of these stars, then SGR
1806-20 is much older than the $\sim 10^3 - 10^4$ yr typically
considered for magnetars.  Alternately, SGR 1806-20 may simply be
taken as evidence for prior massive star formation at this location.
While at least one supernova occurred here, perhaps it was not the
first, and an earlier supernova event triggered the formation of the
currently observed massive stars.

\section{Conclusions}

	We have presented X-ray and infrared observations of the field
of SGR 1806-20, determining the location of the X-ray counterpart in
the IR image to an accuracy of $0.7 \arcsec$ (90\% confidence).  We
find two potential counterparts in this small error circle, and the
crowding in this field makes it difficult to ascertain which if either
of these stars is a true counterpart to SGR 1806-20.  The X-ray
absorption towards SGR 1806-20 matches the infrared extinction of
cluster members, indicating that the SGR 1806-20 is at the same
distance as the cluster ($14.5 \pm 1.4$ kpc), and the IR observations
reveal that it lies within the radial extent of the cluster on the
sky.  The presence of a neutron star in a cluster containing extremely
massive evolved stars indicates either that the neutron star
progenitor was more massive (which seems unphysical) or that these
stars are the result of the most recent of multiple star formation
epochs at this location.

\acknowledgements The authors thank R. Blum and A. Alvarez for help in
obtaining the observations at CTIO.  SSE is supported in part at
Cornell by an NSF CAREER award (NSF-9983830), and SGP and DH were
partially supported by this grant.  MAJ was supported at Cornell by a
NASA Space Grant summer research fellowship, and MAG was supported by a
NSF REU fellowship.

\begin{deluxetable}{lccc}
\tablecaption{Photometry of Stars} 
\startdata 
Star & J-band ($1.25\mu$m) & H-band ($1.65 \mu$m) & K-band ($2.2 \mu$m) \\ 
\hline 
A & $>21$ & $19.5 \pm 1.0$ & $16.7 \pm 0.2$ \\ 
B & $>21$ & $>xx$ & $18.6 \pm 1.0$ \\
C & $19.2 \pm 0.2$  & $17.2 \pm 0.2$ & $16.1 \pm 0.2$ \\
D & $18.8 \pm 0.2$ & $15.5 \pm 0.2$ & $13.87 \pm 0.15$ \\
E & $17.1 \pm 0.2$ & $14.00 \pm 0.15$ & $12.00 \pm 0.11$ \\

\enddata
\end{deluxetable}

\vfill \eject

\begin{figure}
\plotone{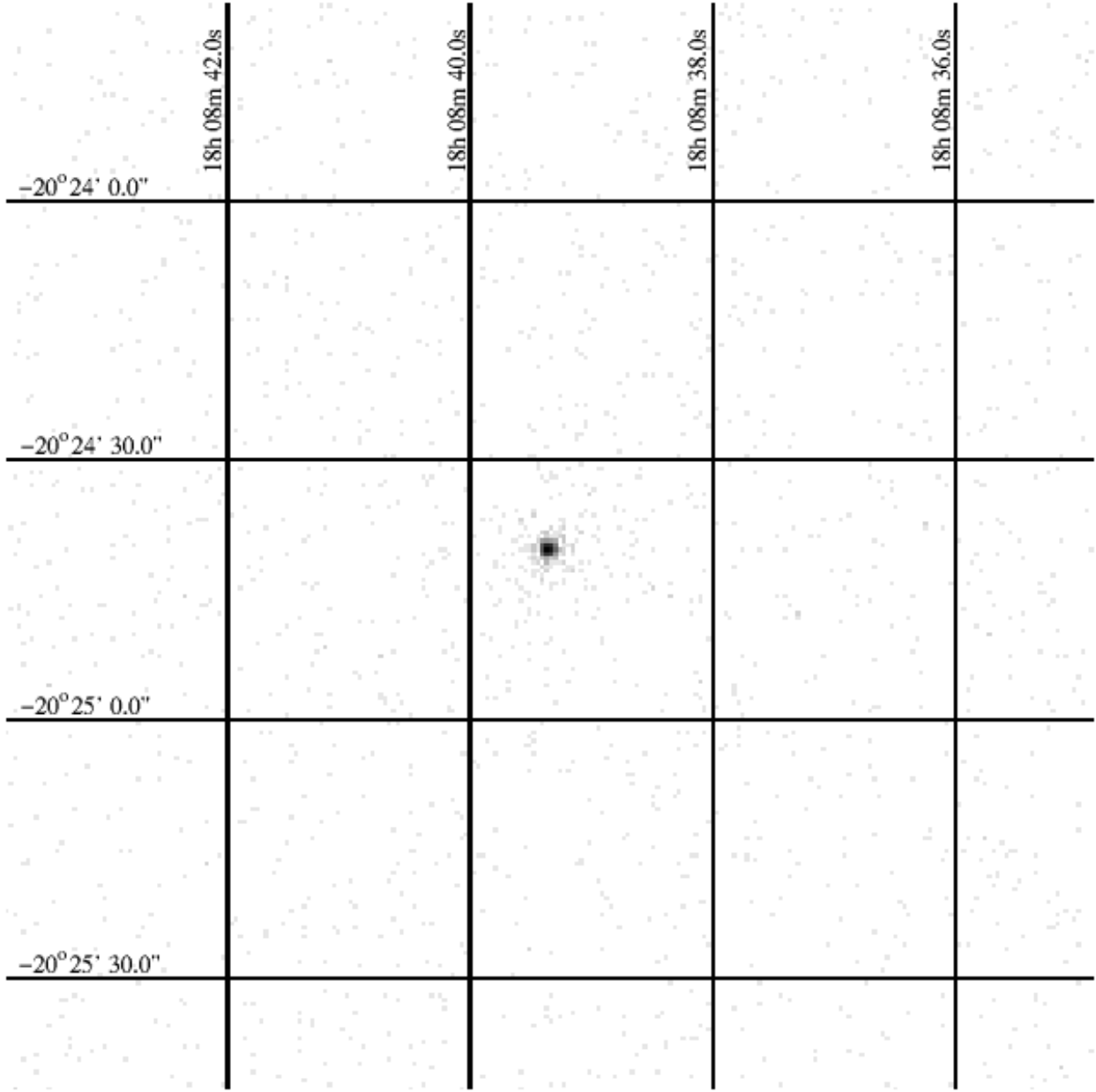}
\caption{\it Chandra ACIS image of the field of SGR 1806-20.  The
field of view is approximately $2 \arcmin$ on a side.  The image
intensity is scaled logarithmically.  Coordinates are J2000.}
\end{figure}

\begin{figure}
\plottwo{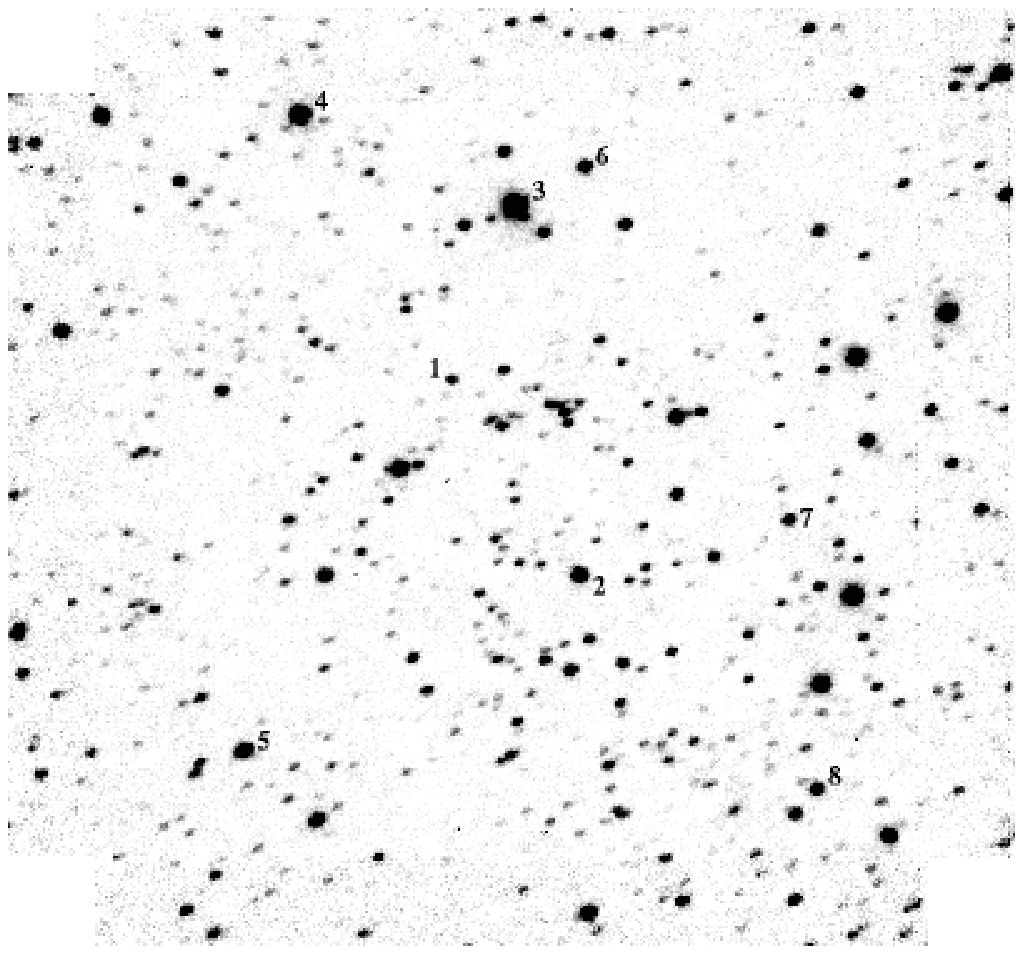}{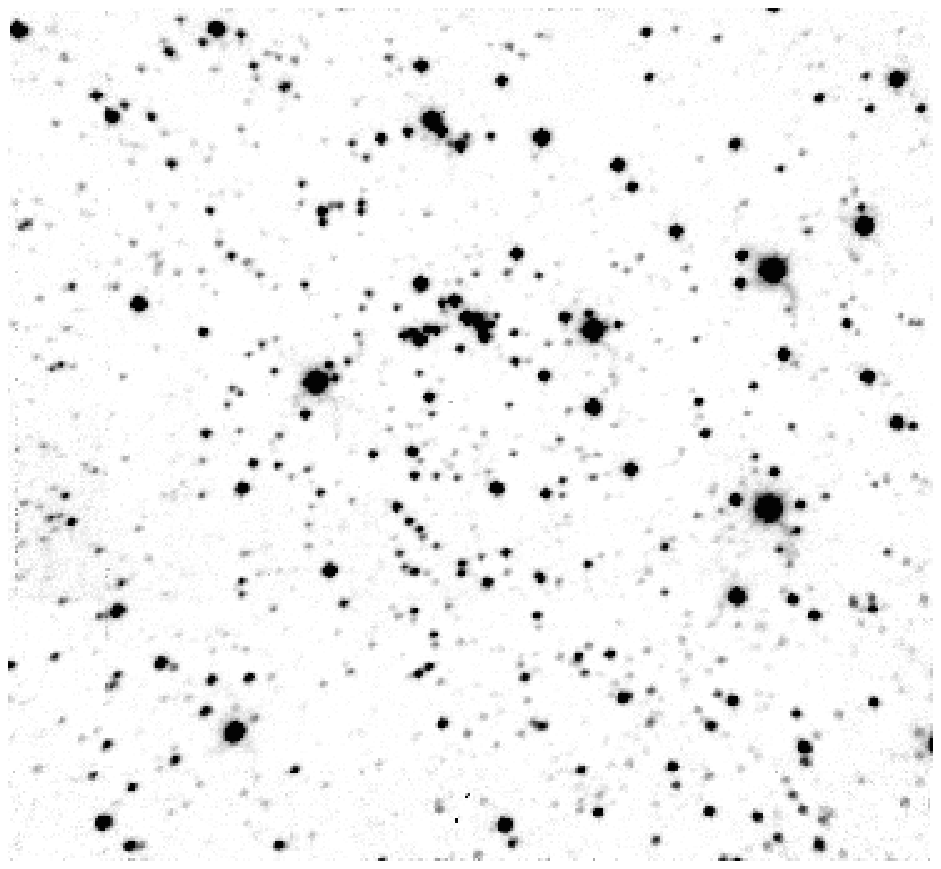}
\plotone{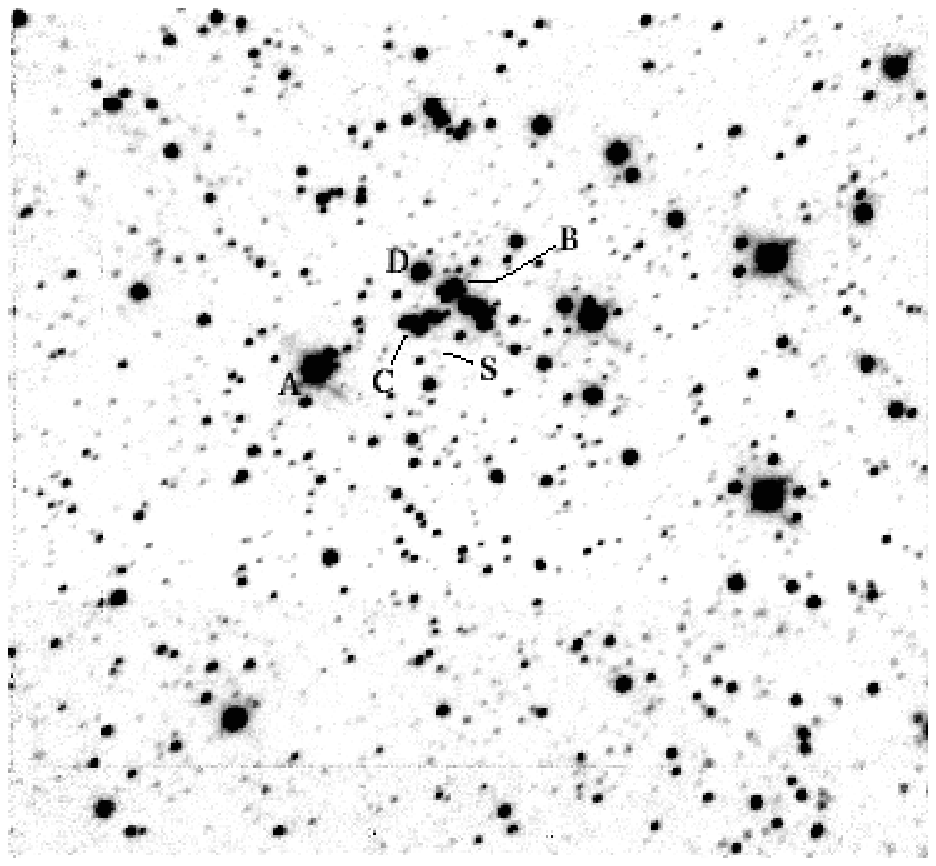}
\caption{\it Near-infrared images of the field of SGR 1806-20 in the
J-band (top left), H-band (top right), and K-band (bottom).  The field
of view is approximately $100 \arcsec$ on a side, with North up and
East to the left.  In the J-band image, numbers indicate the 8
USNO-A2.0 stars used for astrometric calibration of the images.  In
the K-band image, letters indicate the position of SGR 1806-20 (S), as
well as the 4 brightest stars in the nearby cluster (see \citet{LBV}
for details): the luminous star LBV 1806-20 (A), a Wolf-Rayet WCL star
(B), two blue hypergiants of luminosity class Ia+ (C,D).}
\end{figure}

\begin{figure}
{\plottwo{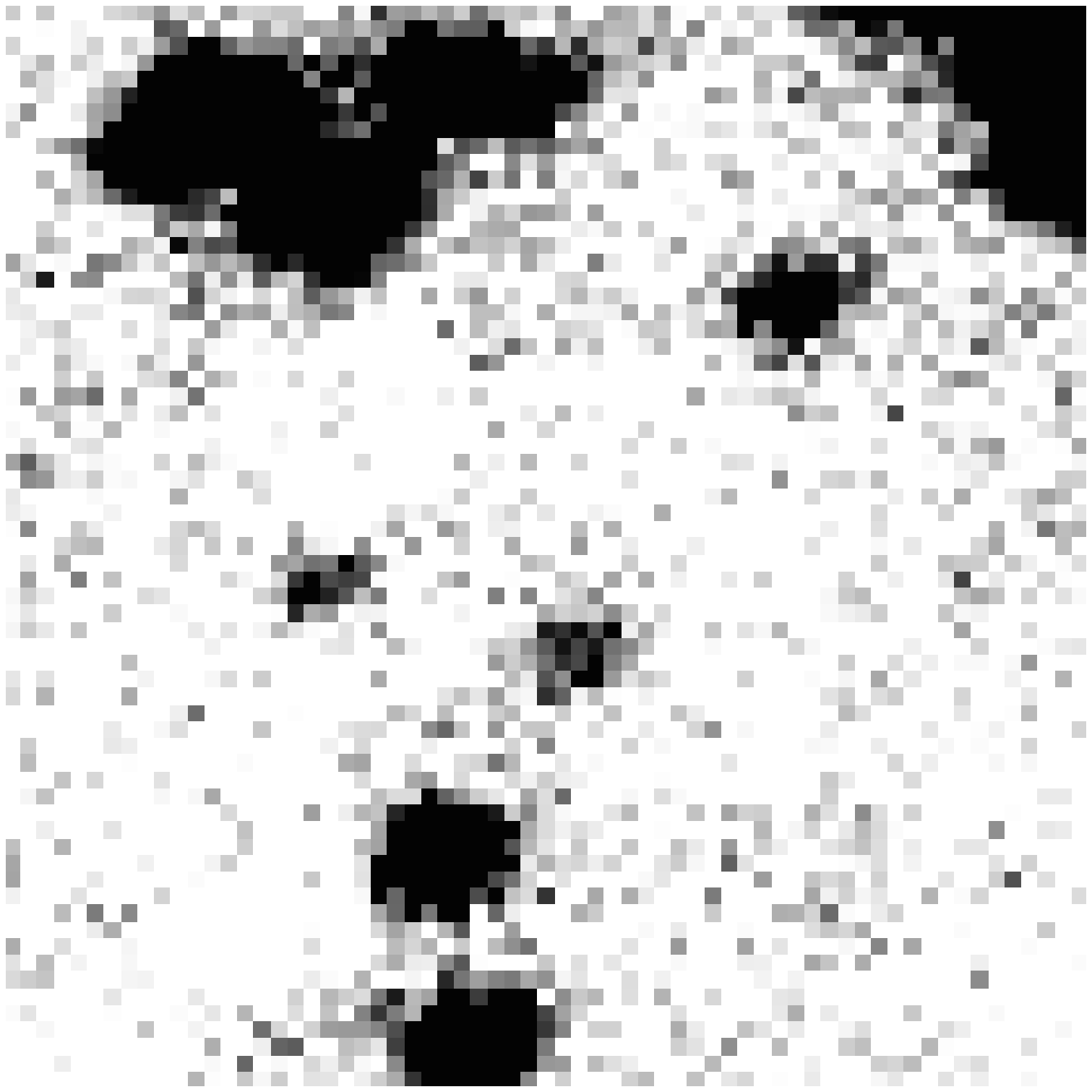}{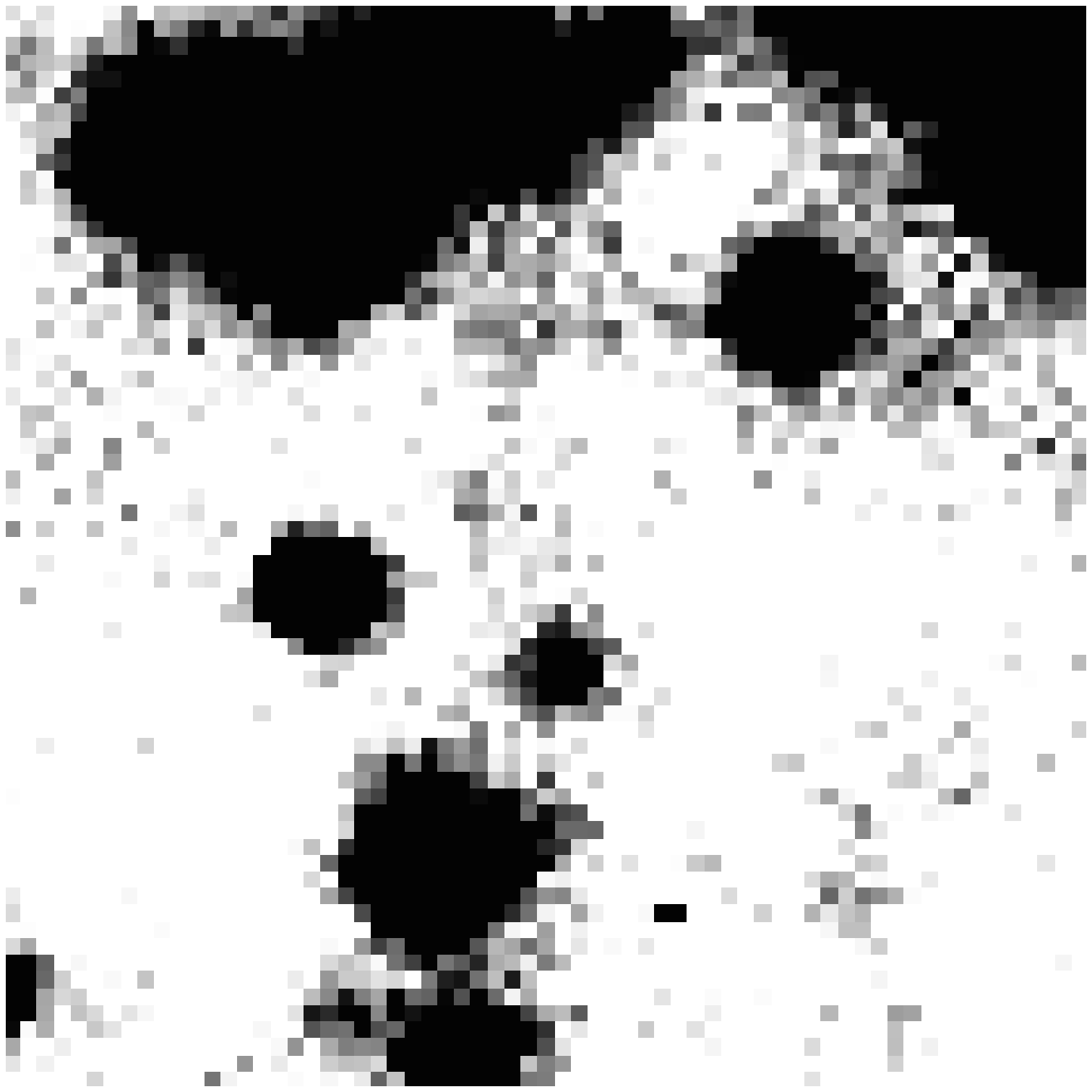}}
\vskip 2.0 mm
{\plottwo{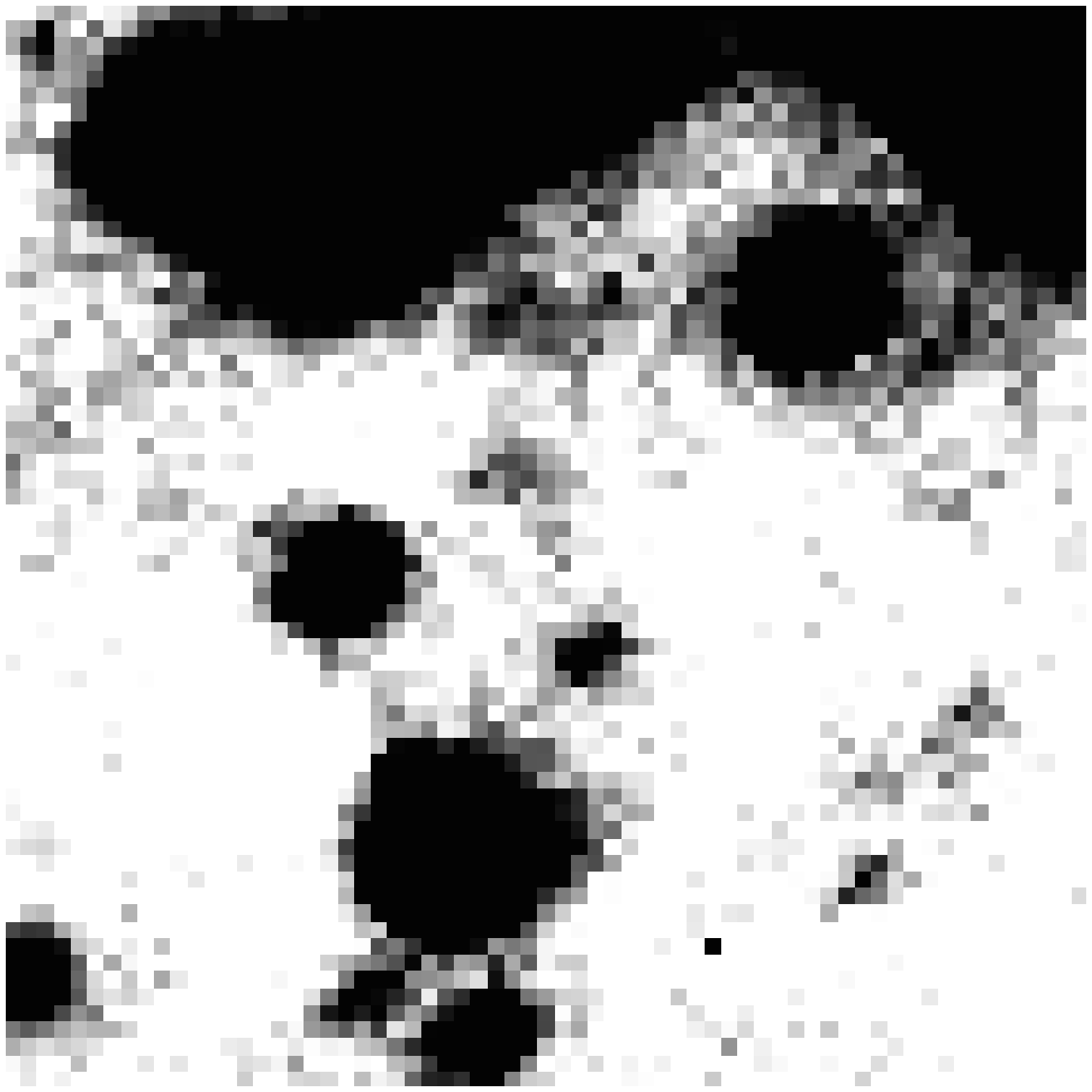}{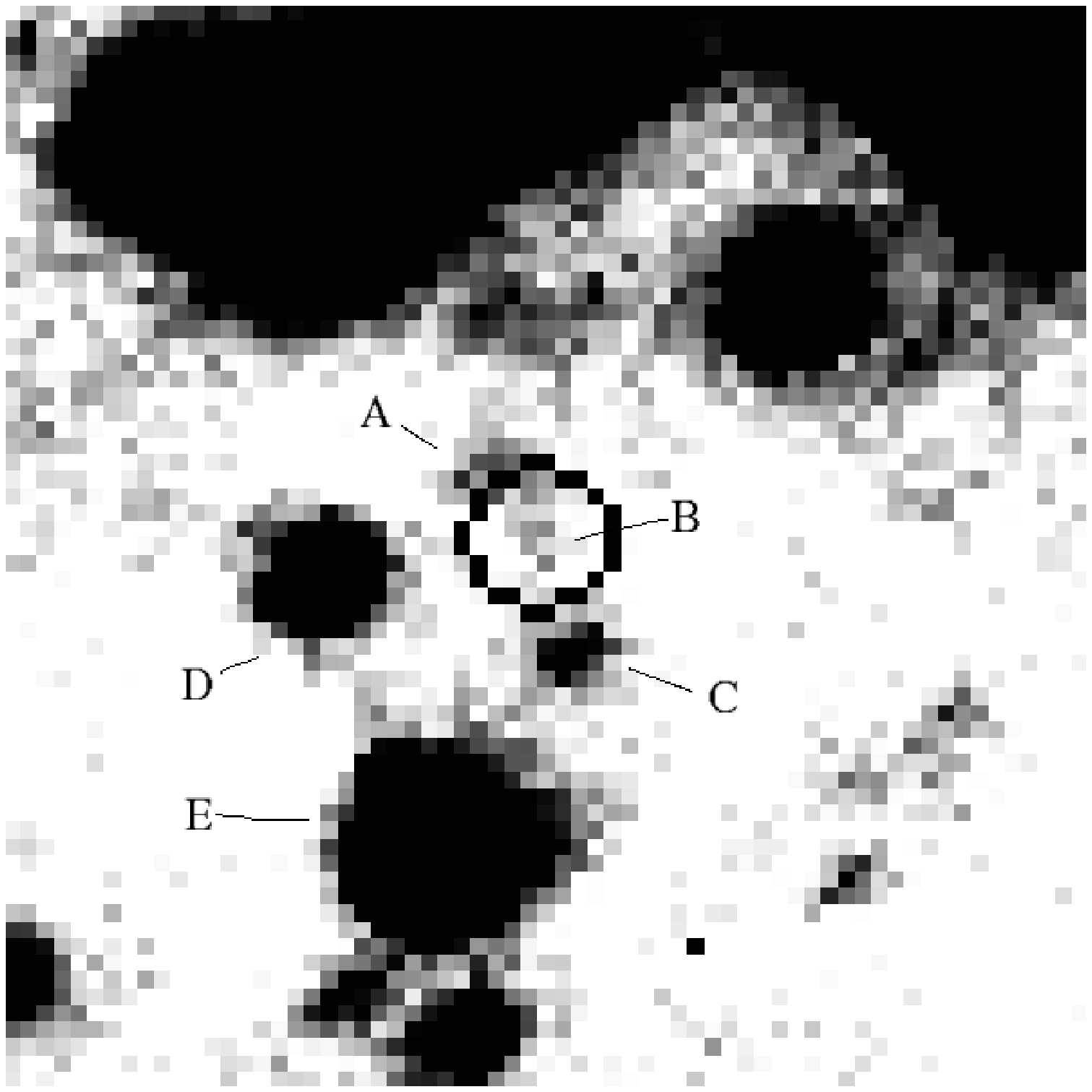}}
\caption{\it Close-up of the region near SGR 1806-20 in the J-band
(top left), H-band (top right), and K-band (bottom left).  The bottom
right image is a copy of the bottom left with the $0.7 \arcsec$-radius
error circle for SGR 1806-20 superimposed, as well labels for the
stars listed in Table 1.  The images are approximately $10 \arcsec$ on
a side.}
\end{figure}

\end{document}